\documentclass[ ]{elsarticle}
\usepackage[margin=2cm,bottom=2cm]{geometry}
\usepackage{chemformula} 
\usepackage[T1]{fontenc} 
\usepackage[export]{adjustbox}
\usepackage{mathptmx,cuted}
\usepackage{physics}
\usepackage{sectsty}
\usepackage{graphicx}
\usepackage{float}
\usepackage{fancyhdr}
\usepackage{fnpos}
\usepackage{array}
\usepackage{droidsans}
\usepackage{charter}
\usepackage[T1]{fontenc}
\usepackage{epstopdf}
\usepackage{esdiff}
\newcounter{qcounter}
\usepackage{amssymb,amsmath,amsthm}
\usepackage{adjustbox,lipsum} 
\usepackage{url}
\usepackage{xr-hyper}
\let\oldsqrt\sqrt
\def\sqrt{\mathpalette\DHLhksqrt}
\def\DHLhksqrt#1#2{%
	\setbox0=\hbox{$#1\oldsqrt{#2\,}$}\dimen0=\ht0
	\advance\dimen0-0.2\ht0
	\setbox2=\hbox{\vrule height\ht0 depth -\dimen0}%
	{\box0\lower0.4pt\box2}}
\makeatletter
\newcolumntype{P}[1]{>{\centering\arraybackslash}p{#1}}
\newsavebox\myboxA
\newsavebox\myboxB
\newlength\mylenA
\newcommand*\xoverline[2][0.75]{%
	\sbox{\myboxA}{$\m@th#2$}%
	\setbox\myboxB\null
	\ht\myboxB=\ht\myboxA%
	\dp\myboxB=\dp\myboxA%
	\wd\myboxB=#1\wd\myboxA
	\sbox\myboxB{$\m@th\overline{\copy\myboxB}$}
	\setlength\mylenA{\the\wd\myboxA}
	\addtolength\mylenA{-\the\wd\myboxB}%
	\ifdim\wd\myboxB<\wd\myboxA%
	\rlap{\hskip 0.5\mylenA\usebox\myboxB}{\usebox\myboxA}%
	\else
	\hskip -0.5\mylenA\rlap{\usebox\myboxA}{\hskip 0.5\mylenA\usebox\myboxB}%
	\fi}
\makeatother

\allowdisplaybreaks  

\begin{document}
	\biboptions{numbers,sort&compress}
\begin{frontmatter}
\title{Evaporating droplets on inclined plant leaves and synthetic surfaces:\\ experiments and mathematical models}	
\author[Scott]{Eloise~C.~Tredenick}
\author[Alison]{W.~Alison~Forster}
\author[Scott]{Ravindra~Pethiyagoda}
\author[Alison]{Rebecca~M.~van~Leeuwen}
\author[Scott]{Scott~W.~McCue}
\ead{scott.mccue@qut.edu.au}

\address[Scott]{School of Mathematical Sciences, Queensland University of Technology, QLD 4001, Australia}
\address[Alison]{Plant Protection Chemistry NZ Ltd., PO Box 6282, Rotorua, New Zealand.}

\begin{abstract}
{\em Hypothesis}\newline
Evaporation of surfactant droplets on leaves is complicated due to the complex physical and chemical properties of the leaf surfaces.  However, for certain leaf surfaces for which the evaporation process appears to follow the standard constant-contact-radius or constant-contact-angle modes, it should be possible to mimic the droplet evaporation with both a well-chosen synthetic surface and a relatively simple mathematical model.

\vspace{0.5ex}\noindent
{\em Experiments}\newline
Surfactant droplet evaporation experiments were performed on two commercial crop species, wheat and capsicum, along with two synthetic surfaces, up to a $90\,^{\circ}$ incline.  The time-dependence of the droplets' contact angles, height, volume and contact radius was measured throughout the evaporation experiments.  Mathematical models were developed to simulate the experiments.

\vspace{0.5ex}\noindent
{\em Findings}\newline
With one clear exception, for all combinations of surfaces, surfactant concentrations and angles, the experiments appear to follow the standard evaporation modes and are well described by the mathematical models (modified Popov and Young-Laplace-Popov).  The exception is wheat with a high surfactant concentration, for which droplet evaporation appears nonstandard and deviates from the diffusion limited models, perhaps due to additional mechanisms such as the adsorption of surfactant, stomatal density or an elongated shape in the direction of the grooves in the wheat surface.

\end{abstract}

\end{frontmatter}

\section{Introduction}\label{sec:lit}
Increasing the efficacy of agrochemicals can reduce the usage of water and active ingredients, minimise environmental impacts and maximise crop yield.  Spray droplet evaporation time has been found to have a significant effect on the penetration of active ingredients through plant cuticles \cite{Ramsey2005review} and therefore biological efficacy. Here we report on an experimental and mathematical study of evaporation of surfactant droplets on both plant and synthetic surfaces with a view to better understand these processes from a fundamental perspective and in the context of agrochemical spraying of plants.

We focus on two plant species, one graminaceous (wheat, \textit{Triticum aestivum} L.) and one broad-leaf (capsicum, \textit{Capsicum annuum} L.,~variety Giant Bell), having extremely different canopy architectures (vertical vs relatively horizontal), as well as different wettabilities (very difficult-to-wet vs moderate).
Wheat is of major agricultural importance world-wide and efficient foliar fertilisers are required to ensure adequate nutrient delivery; foliar application (application to plant leaves) being advantageous during growing seasons when climatic conditions are known \cite{Peirce2016}.  We consider a range of inclination angles, as plant leaf inclination angles vary widely, not only across species (from horizontal broad-leaf species through to vertically oriented graminaceous species) but also within the same species, and this is particularly true for wheat leaves.  We compare evaporation of droplets on wheat and capsicum leaves with that on Teflon and Parafilm.  The synthetic surfaces have similar surface characteristics to the plant leaves and one of the aims is to determine to what extent they are able to replicate droplet evaporation on the plant leaves.  Agrochemical spray droplets often include adjuvants such as surfactants which reduce the surface tension, leading to a reduced initial contact angle and improved wetting \cite{Gimenes2013}. In this study, the wetting agent Surfynol 465 (S465) is chosen as it reaches a low equilibrium surface tension rapidly.

Evaporation of a sessile droplet on a substrate is a topic of significant interest \cite{Picknett1977,Bourges1995,Cachile2002,Erbil2002,Hu2002,McHale2005}
whose vast literature has been summarised in various review articles \cite{Cazabat2010,Erbil2012,Kovalchuk2014,Brutin2018}.
The process is heavily influenced by the wettability and roughness of the surface, with the key measurements being the contact radius $R(t)$, contact angle $\theta(t)$ and height of the droplet $H(t)$.  Under ideal circumstances, the phases of a partial wetting evaporating sessile drop are as follows \cite{Picknett1977,Dash2013,Semenov2013,Kovalchuk2014,Semenov2014,Stauber2014,Brutin2018} (see also Fig.~S1 in the Supplementary Material):
\begin{list}{\arabic{qcounter})~}{\usecounter{qcounter}}
	
	\item \underline{Spreading -- S}: A droplet is initially deposited on a surface and quickly spreads with a decreasing contact angle and increasing contact radius until an initial contact angle ($\theta_0$) and radius ($R_{0}$) are reached. This process is very fast and may occur in under two minutes for adjuvant solutions on leaf surfaces \cite{Xu2011}. Sometimes evaporation is neglected here due to the relatively short time scale.
	
	\item \underline{Stage 1 -- CCR}: The first stage of evaporation, referred to as the constant contact radius (CCR) mode, is characterised by a decreasing contact angle ($\theta(t)$) and a constant contact radius ($R_{0}$). The contact angle changes between the initial contact angle ($\theta_0$) and the receding contact angle ($\theta_{\text{rec}}$).
	
	\item \underline{Stage 2 -- CCA}: The second stage of evaporation, referred to as the constant contact angle (CCA) mode, is characterised by a constant contact angle ($\theta_{\text{rec}}$) and a decreasing contact radius ($R(t)$).
	
	\item \underline{Extinction}: The final phase of evaporation occurs over a short time and continues until the droplet completely evaporates to extinction. It is characterised by a changing contact angle and radius.
\end{list}
In practice, the evaporation of a sessile droplet may not be so straightforward.  For example, while the CCR mode is commonly observed, often the CCA mode only holds approximately \cite{Cazabat2010}.  Another possibility in practice is that evaporation proceeds under a ``stick-slip'' regime, where part of the process occurs under the CCR mode and then the CCA mode, but returns to the CCR mode and then again the CCA mode, and so on \cite{Shanahan1995,Cazabat2010}.  Finally, it is possible that (perhaps following CCR and CCA modes) both the contact angle and contact radius decrease in time together in a mixed or combined mode \cite{Picknett1977}. We note the receding time, $t_{\text{rec}}$, which is the time when the evaporation modes switch from CCR to CCA mode, may vary significantly \cite{Xu2013Micro} and many examples do not include a CCA mode.

Droplet evaporation on plant surfaces is potentially much more complex than on synthetic surfaces and has only received moderate attention in the literature \cite{Gimenes2013,Januszkiewicz2019effect,Peirce2016,Tredenick2018,Xu2011,Zhou2017,Yu2009c,Xu2010,Zhou2018,Zhang2006evaporation}.
Given the commonly applied context of agrochemical spraying, the effect of surfactant and leaf surface properties are typically the focus.  Regarding experiments with wheat leaves, droplets of various concentrations and solutions have been applied to adaxial and abaxial wheat leaves, and concentration and type of adjuvant had a significant impact on the initial droplet contact angle and adsorption of adjuvant \cite{Zhang2017Soft,Januszkiewicz2019effect,Zhang2006}.  In an experimental study by Peirce et al.~\cite{Peirce2016} regarding uptake and evaporation on wheat leaves with various surfactant solutions, the authors find the contact angle quickly decreases with time and, in one case, reduces to below 10$\,^{\circ}$ in around 10~sec, while the other two surfactants take 10 and 15~min. They describe the results as a type of stick-slip regime, where the contact line is temporarily pinned until a condition for de-pinning is reached.

There are several mathematical models present in the literature \cite{Deegan2000,Nguyen2012,Stauber2015evaporation,Dunn2009,Sefiane2009,Stauber2014,Dunn2008,Talbot2012} that consider different aspects of droplets on an incline: a simple model to calculate (but not predict) the volume of a droplet on an incline \cite{ElSherbini2004}; using energy-minimising techniques to model the static droplet shape on an incline \cite{Merte1987,Milinazzo1988,Dimitrakopoilos1999,Iliev1997}; model the shape of the droplet on an incline as a function of volume, contact radius and incline angle \cite{Brown1980}; and model the droplet evaporation using a variation of Laplace's equation with flexible geometries that could be extended to droplets on an incline \cite{Saenz2017dynamics}.

There are three aims of this study.  First, we conduct new experiments on droplet evaporation on leaf surfaces with a view to clarifying the effect that surfactant concentration and angle of surface inclination has on the mode of evaporation and other relevant outputs such as droplet shape, contact angle, evaporation rate and evaporation time. Second, we investigate the extent to which evaporation of surfactant droplets on horizontal and inclined leaf surfaces is analogous to that on synthetic surfaces with similar surface properties.  Finally, we present new relatively simple mathematical models based on the commonly adopted diffusion-limited approach to simulate the experiments.  These mathematical models allow the incorporation of two different contact angles, two evaporation modes (CCR and CCA) and predict changes in the angles, volume, radius and height with time. We apply the models to our new experimental results of water and S465 droplets on wheat, Teflon, capsicum and Parafilm on $0\,^{\circ}$, $45\,^{\circ}$ and $90\,^{\circ}$ inclines.

\section{Experimental Methods}\label{sec:experiment}

\subsection{Surfaces}\label{Exp_Surfaces}
\begin{figure} [h!]
	\centering
	\includegraphics[height=0.4\textheight,keepaspectratio]{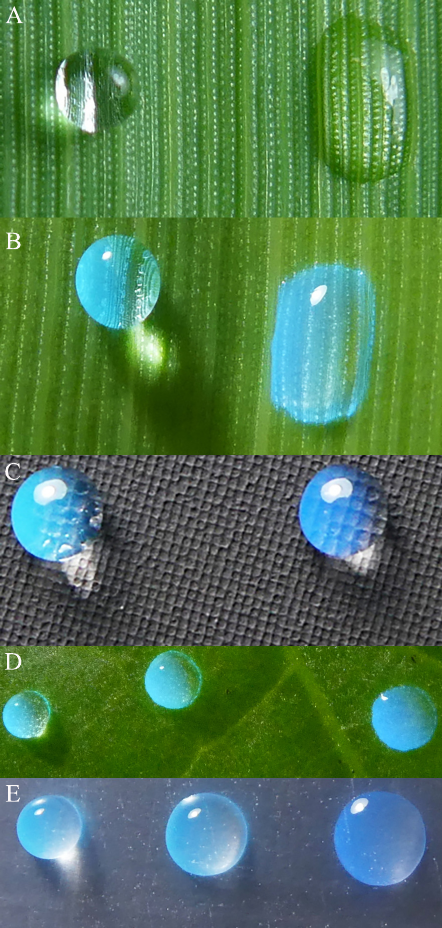}
	\caption{Droplets on wheat (A,B) and capsicum (D) upper leaf surfaces, Teflon (C) and Parafilm (E). The right  droplet is 1\% S465, then 0.1\% S465 and water in (D) and (E). The surfaces are all on a 90$\,^{\circ}$ incline with each 1~$\mu$L droplet's profile captured very soon after placed on the surface. In (B) to (E) the droplets contain 0.5\% Blankophor P 167\% fluorescent dye (Bayer AG, Leverkusen, Germany) and were photographed under UV light with the addition of some white (normal) light, while in (A) no dye is added. Images are not to scale and the scale is different in each subfigure.  Note the CAM apparatus captures its images from the right hand side of the images in this figure, with the wheat leaf ridges perpendicular to the camera.}
	\label{fig:allsurf902}
\end{figure}

Two plant species, capsicum (\textit{Capsicum annuum} L.,~variety Giant Bell) and wheat (\textit{Triticum aestivum} L.), and two synthetic surfaces, rough Teflon coated slides (ES-2013B-CE24 full mask Teflon slide, Erie scientific LLC, Portsmouth NH 03801) and Parafilm (American National Can\textsuperscript{TM}, Chicago, IL 60631, USA) were studied. The plants were grown from seed in individual pots containing PPC\textsubscript{NZ}/Bloom potting mix (Daltons, Hinuera Rd West, PO Box 397, Matamata, NZ), under controlled environment (CE) conditions with 70\% RH, 24 hour water cycle, 12 hour photoperiod at ca.~450~$\mu$mol~m$^{-2}$s$^{-1}$, day/night temperatures $23\,^{\circ}\mathrm{C}$/$15\,^{\circ}\mathrm{C}$ for capsicum and $20\,^{\circ}\mathrm{C}$/$15\,^{\circ}\mathrm{C}$ for wheat. The capsicum and wheat plants were used at approximately seven and six weeks of age, respectively. The youngest, fully expanded, visually healthy leaves were selected. The wheat and rough Teflon surface are both very difficult-to-wet (droplets of 20\% acetone are completely repulsed \cite{Gaskin2005}), both surfaces being non-polar and very rough \cite{Nairn2011quantification}, although the rough Teflon is much rougher and slightly more non-polar than the wheat. The capsicum surface is moderate-to-wet (20\% acetone contact angle of 75$\,^{\circ}$), being slightly non-polar but very smooth, while the Parafilm surface is difficult-to-wet (20\% acetone contact angle of 103$\,^{\circ}$), being similarly smooth as capsicum but much more non-polar (Parafilm is the most non-polar surface of the 4 surfaces studied). Our rough Teflon is a specifically manufactured synthetic superhydrophobic surface with distinct characteristics to mimic the surface of a superhydrophobic plant leaf. The structure of the Teflon surface is comprised of circular valleys and pointed pinnacles, as shown in Fig.~\ref{fig:allsurf902}(C).

\subsection{Surfactant}\label{Exp_Surfactant}
Water (deionised, distilled) was used alone or in the presence of S465 (0.1\% and 1\%; Ethoxylated 2,4,7,9-tetramethyl 5 decyn-4,7-diol, supplied by Syngenta U.K., batch 914432). Droplets of pure water were repulsed by the very difficult-to-wet wheat surface and rough Teflon coated slide surface \cite{Huet20}, and therefore measurements of evaporation on these surfaces could only be taken for the two concentrations of surfactant and not pure water. All three solutions (pure water, 0.1\% and 1\% S465) were applied to both the moderate-to-wet capsicum and difficult-to-wet Parafilm surface. The concentrations were chosen to represent a wide range of contact angles on the surfaces. The equilibrium surface tensions for pure water, 0.1\% and 1\% S465 are 72.8~mN$/$m, 42.4~mN$/$m and 26.7~mN$/$m, respectively.

\subsection{Evaporating droplet experiments}\label{Exp_ContactAngle}
\begin{figure} [h!]
	\centering
	\includegraphics[width=0.45\textheight]{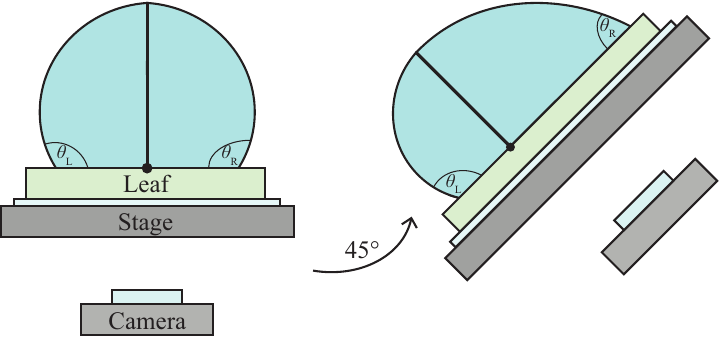}
	\caption{Experimental schematic. The sample and camera are mounted on the tilting stage and move in unison. On the left, the droplet on a horizontal surface is mounted on a glass slide on the stage with the camera. Then the stage can be tilted, for example in the right image to 45$\,^{\circ}$. When the surface is inclined, $\theta_{\text{L}}$ is the inferior angle (closer to the ground), while $\theta_{\text{R}}$ is the superior contact angle.}
	\label{fig:setup2}
\end{figure}
Measurements of left and right contact angle, droplet height, volume and diameter were taken throughout the evaporation experiment, from immediately after the 1~$\mu$L droplets had been placed on the leaf (adaxial/top) or synthetic surface until the droplet had evaporated (see Fig.~S2 in the Supplementary Material for a schematic). Contact angles were measured using an optical contact angle meter with an automated tilting stage (CAM 200; KSV Instruments, Helsinki, Finland) with a digital camera (Basler AG.~Ahrensburg, Germany). The tilting stage was set to either horizontal, 45$\,^{\circ}$ or 90$\,^{\circ}$ for the duration of the evaporation experiment. The left side was angled downwards and the camera tilted with the stage.  The angle $\theta_{\text{L}}$ is the larger contact angle that is most deformed by gravity on an incline; it is therefore the inferior angle (closer to the ground), while $\theta_{\text{R}}$ is the superior contact angle (Fig.~\ref{fig:setup2}).  For each image taken, droplet volume was estimated by the CAM software (version CAM4.0.1) (except for three cases in which we used the droplet height and contact angles to estimate the volume using a spherical cap model, equation (S1) in the Supplementary Material).  The wheat leaves were positioned so the ridges were perpendicular to the tilting stage and camera. Examples of images taken under these experimental conditions can be seen in Fig.~\ref{fig:alldrops1}, while Fig.~\ref{fig:allsurf902} was not taken with the tilting stage camera. Five replicates were performed per treatment and the results were averaged. The experiments were performed in a temperature range of $18\,^{\circ}\mathrm{C}$ - $20\,^{\circ}\mathrm{C}$, with 50\% - 52.5\% RH. The measurements are taken every 0.5 sec for the first 30 sec, then every 30 sec. The volume was set to zero when the droplet had completely evaporated by observing the images captured during the experiment.

\section{Model Framework} \label{sec:model}

We consider two mathematical models based on the assumption of quasi-steady diffusion-limited evaporation.  The first we refer to as the \citet{Popov2005} model, which can simulate droplet evaporation under either CCR or CCA modes on a horizontal substrate for the case in which the droplet shape is considered to be a spherical cap.  When we apply this model on an inclined substrate, we average the initial contact angles and refer to the modified version as Popov averaged (Popov Av).  The second is our Young-Laplace-Popov model, which is an extension of the Popov model that allows the geometry of the droplet to deform from a spherical cap according to the effects of gravity and surface tension. (Table S2 describes the parameters considered over the models presented here.)  Finally, we note that in the Supplementary Material we also include a third model that we have developed to apply on an incline, again under either CCR or CCA modes, by incorporating two half-spherical caps with two different contact angles with the option of a non-circular contact line.  We call this the droplet-incline-Popov model (DIP model).

\subsection{Popov model} \label{sec:Popov}
To account for the evaporation of a sessile droplet of water and surfactant, we will assume evaporation occurs as the sequence of CCR mode followed by CCA mode \cite{Picknett1977}.  As such, we are able to utilise the quasi-steady diffusion-limited evaporation model for a spherical cap-shaped droplet, which we are going to call the Popov model, since \citet{Popov2005} wrote out the full details (although the actual model is treated by others \cite{Picknett1977,Deegan2000,Erbil2012,Dash2013} and is equivalent to an older electrostatics problem). In the context of evaporation on plant surfaces, the Popov model has been applied with surfactant solutions on rice \cite{Zhou2017}, which has a similar contact angle to wheat \cite{he2019influence}, and \citet{Tredenick2018} modified the Popov model to account for hygroscopic water absorption on tomato fruit cuticles with surfactant.

The \citet{Popov2005} model for a horizontal substrate is as follows.  We require as inputs to the model the initial contact angle $\theta_0$ and the initial droplet height $H_0$.  Since we are assuming the droplet takes the shape of a spherical cap, we can then easily use geometry to compute the initial contact radius $R_0$ and the initial volume $V_0$ (alternatively if we measure the initial contact radius $R_0$ instead of the initial droplet height $H_0$, then the other can easily be calculated). The model requires us to specify the receding time when the evaporation modes switch from CCR to CCA, which we call $t_{\text{rec}}$ (other authors \cite{Nguyen2012b} call this the transition contact angle). Finally, we require other experimental inputs to compute the evaporation constant $\Lambda$, such as the diffusivity of water vapour in air, temperature, relative humidity and so on (see Table~S2).  The governing equations for the Popov model can be written as the coupled system of algebraic-differential equations
\begin{equation}
\diff{V}{t} = - \pi \ \Lambda \ R \ f(\theta),
\quad
V=\frac{\pi}{3j(\theta)} R^3,
\label{eq:popovmodel}
\end{equation}
where
\begin{align}
j(\theta) &= \dfrac{\sin^3\theta}  { (1-\cos\theta)^2 \ (2+\cos\theta)}, \label{gtheta2}\\
f(\theta) &= \tan\left(  \dfrac{\theta}{2} \right) + 8 \int\limits_0^{\infty} \cosh^2(\theta u) \ \text{csch}(2  \pi  u) \ \tanh \left[ \left(  \pi - \theta   \right) u \right] \ \mathrm{d} u. \label{ftheta2}
\end{align}
For the period of time in which evaporation is governed by the CCR mode, the radius is fixed to be $R=R_0$ for $0 < t \leq t_{\text{rec}}$.  For $t>t_{\text{rec}}$, the assumption is that the evaporation proceeds in the CCA mode, for which $\theta=\theta_{\text{rec}}$.
The assumptions behind the derivation of (\ref{eq:popovmodel}) are summarised in the Supplementary Material. Note that the Popov model is only designed for horizontal substrates. When we apply it to droplet evaporation on an incline, the initial left and right contact angles are simply averaged to determine an initial contact angle for the model.  We emphasise that this averaging procedure only applies to the original Popov model when we attempt to apply it to evaporation on an incline (for our new model described below, the left, $\theta_{\text{L}}$, and right,  $\theta_{\text{R}}$, contact angles are tracked explicitly).

\subsection{Young-Laplace-Popov model}\label{sec:numericalMethod}
We consider an extension of the \citet{Popov2005} evaporation model whereby the shape of the droplet is no longer considered to be a spherical cap, but is instead determined by the solution to the Young-Laplace equation.  We call this new approach the Young-Laplace-Popov (YLP) model. For simplicity, we have assumed the contact area of the droplet is circular during the evaporation process, even when the droplet is on an incline.  The YLP model can be summarised with a time stepping scheme which, for a given droplet shape from the previous time, involves three steps:
\begin{enumerate}
	\item Solving a three-dimensional steady-state vapour transport problem to compute the water vapour flux at the droplet surface and thus the change in droplet volume due to evaporation.
	\item Updating the droplet volume for one time step.
	\item Computing the new droplet shape by satisfying the Young-Laplace equation subject to the new volume and either a specified constant contact radius (CCR mode) or constant contact angle at the down-plate edge (CCA mode) from the previous time-step.
\end{enumerate}
The initial droplet shape can be computed by specifying any two properties from the contact radius $R_0$, larger contact angle $\theta_{\text{L0}}$, droplet volume $V_0$ or droplet height $H_0$, and then satisfying the Young-Laplace equation. Note that the two properties not initially specified will be determined as part of the solution to the Young-Laplace equation. As well as the experimental parameters used in the original Popov model, our YLP model requires as an input the value of surface tension.

The volume of liquid evaporated at each time-step is determined by solving the steady-state transport problem governed by Laplace's equation,
\begin{equation}
\nabla^2 \ c = 0 \label{Laplace1},
\end{equation}
where $c$ is the water vapour concentration. The domain has four boundaries: the exterior droplet surface $\partial V_D$, the synthetic or plant leaf surface $\partial V_P$, a symmetry plane $\partial V_S$ and the atmosphere $\partial V_A$. Due to the symmetry of the problem, the domain is bounded by the substrate, the droplet's exterior surface, a large sphere of air, and a symmetry plane. The domain must be re-meshed at every time step to account for the changing droplet shape. We define the transport domain as a quarter sphere with the shape of the droplet removed from the centre of the sphere. On the droplet surface the air is saturated with vapour, there is no vapour flux through the plate or symmetry plane, and far away from the droplet the air has some atmospheric vapour concentration $c_\infty$. Because the radius of the quarter sphere is not at infinity, we approximate the atmospheric condition with a boundary condition analogous to Newton's law of cooling. Therefore, the boundary conditions can be written as
\begin{equation}
c|_{\partial V_D}=c_s, \quad \left.\frac{\partial c}{\partial n}\right|_{\partial V_P}=0,
\end{equation}
\begin{equation}
\left.\frac{\partial c}{\partial n}\right|_{\partial V_S}=0, \quad \left.\frac{\partial c}{\partial n}\right|_{\partial V_A}=\frac{c_\infty-c|_{\partial V_A}}{r_{\infty}},
\label{eq:BCatmos}
\end{equation}
where $\partial / \partial n$ is the outward normal derivative, $c_s$ is saturated water vapour concentration and $r_\infty$ is the radius of the domain for the quarter sphere of air.

The governing equations (\ref{Laplace1})--(\ref{eq:BCatmos}) are solved using the MATLAB\textsuperscript{\textregistered} PDE toolbox, which uses a finite volume scheme with a quadratic tetrahedral mesh. The total vapour flux at the droplet surface,
\begin{align}
J&=  -D_\mathrm{evap}\iint\limits_{\partial V_D}\frac{\partial c}{\partial n}\,\mathrm{d}S, \label{eq:Jflux}
\end{align}
is computed from the solution and hence the change in droplet volume is determined via
\begin{align}
\diff{V}{t}&=\frac{J}{\Delta\rho}, \label{eq:Vflux}
\end{align}
where $\Delta\rho$ is the density difference between the droplet and air. Further details are provided in the Supplementary Material.

We note that our YLP approach is similar to that used in a very recent study by Timm et al.~\cite{Timm2020} (which was published after our model was developed), who also consider evaporating droplets on an incline.  Like our YLP model, they assume the contact line is circular, solve the Young-Laplace equation to determine the shape of the droplet, and compute the flux of liquid at the air/droplet interface by solving Laplace's equation outside the droplet.  However, in contrast to our approach, in their study they do not evolve the droplet shape in time as the evaporation process continues so they do not calculate quantities such as contact angle, droplet height, contact radius and droplet volume as a function of time.

\section{Results and Discussion} \label{results}
\subsection{Experiments on wheat and Teflon}  \label{sec:expwheatteflon}

\begin{figure*}  [h!]
	\includegraphics[width=0.8\textwidth, center,keepaspectratio]{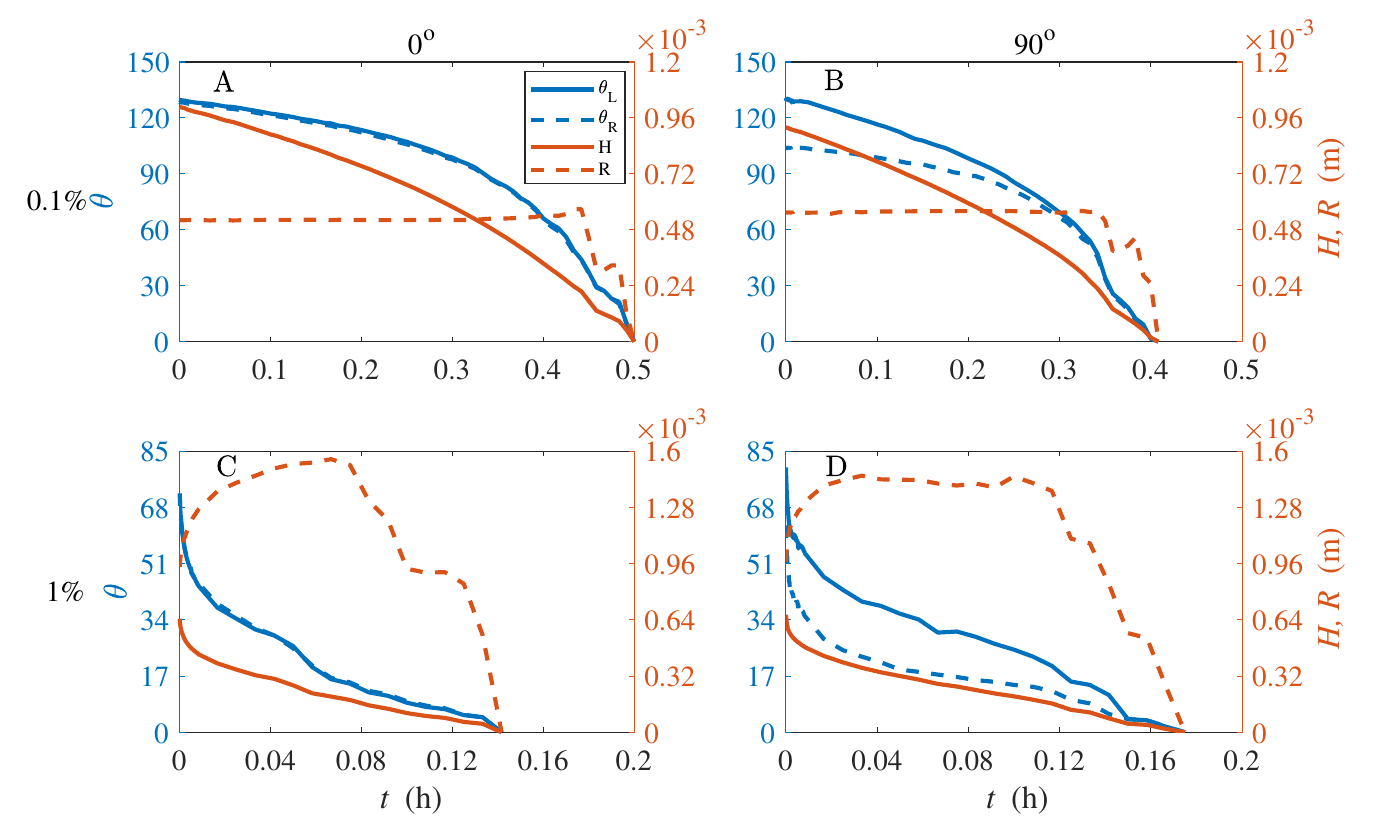}
	\caption{Experimental data for droplet evaporation on wheat leaves. 0.1\% S465 is shown in A and B, and 1\% S465 is shown in C and D. The columns of plots correspond to 0$\,^{\circ}$ and 90$\,^{\circ}$ surface incline. The left and right contact angles, $\theta_L$ and $\theta_R$, are in blue, corresponding to the left hand side axis, while the contact radius $R$ and height $H$ are both in red and correspond to the right hand side axis.  The horizontal axis measures time, $t$.  Note the time scale in (A)--(B) is different to (C)--(D).}
	\label{fig:wheatallsubs1}
\end{figure*}

\begin{figure*}  [h!]
	\includegraphics[width=0.8\textwidth, center,keepaspectratio]{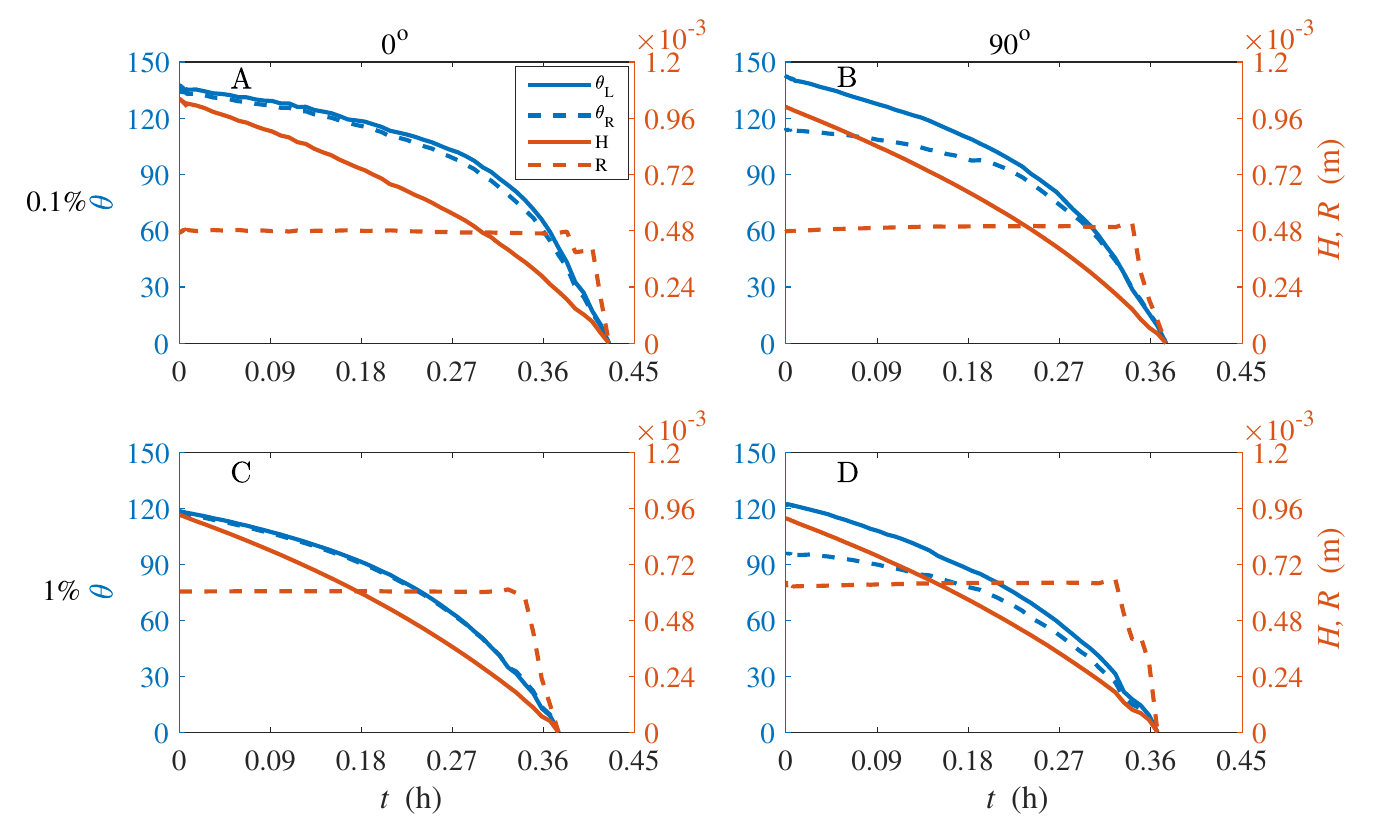}
	\caption{Experimental data for droplet evaporation on Teflon. 0.1\% S465 is shown in A and B, and 1\% S465 is shown in C and D. The columns of plots correspond to 0$\,^{\circ}$ and 90$\,^{\circ}$ surface incline. The left and right contact angles, $\theta_L$ and $\theta_R$, are in blue, corresponding to the left hand side axis, while the contact radius $R$ and height $H$ are both in red and correspond to the right hand side axis.  The horizontal axis measures time, $t$.}
	\label{fig:teflonallsubs1}
\end{figure*}

We begin with the experimental results for wheat and Teflon surfaces (Fig.~\ref{fig:wheatallsubs1}--\ref{fig:teflonallsubs1}). Each plot in these figures illustrates the dependence of the left contact angle $\theta_{\text{L}}$, the right contact angle $\theta_{\text{R}}$, the contact radius $R$ and droplet height $H$ on time.  Recalling that pure water droplets do not adhere on either wheat or Teflon due to their very difficult-to-wet surfaces, we could not run the experiments for water; as such, these figures show only results for 0.1\% (top row in each figure) and 1\% S465 (bottom row).  In each case we have plots for horizontal surfaces (first column) and surfaces inclined at 90$\,^{\circ}$ (second column).  Note the results for surfaces inclined at 45$\,^{\circ}$ are show in the Supplementary Material.

The most obvious observation from these figures is that, for low concentration of surfactant (0.1\% S465), the contact radius $R$ appears fixed in time (except for the final moments when the droplet volume is low) and so the evaporation mode is CCR for both wheat and Teflon for essentially the entire lifetime of the process (top row of Fig.~\ref{fig:wheatallsubs1}--\ref{fig:teflonallsubs1}).  As expected, the difference between the left and right initial contact angles, $\theta_{\text{L}}-\theta_{\text{R}}$, increases with increasing substrate inclination angle; the largest difference between these contact angles is 28.5$\,^{\circ}$ for 0.1\% Teflon on a 90$\,^{\circ}$ incline, as shown in Fig.~\ref{fig:teflonallsubs1} (B).  On an incline, $\theta_{\text{L}}-\theta_{\text{R}}$ decreases as time progresses as the effect of gravity on the droplet shape diminishes due to its smaller size.  The continued CCR mode for the effective lifetime of these evaporating droplets, even on a 90$\,^{\circ}$ incline, demonstrates a strong tendency for the contact line to remain pinned on these very difficult-to-wet surfaces (similar observations have recently been made for evaporating droplets on so-called biphobic surfaces at 90$\,^{\circ}$ inclines \cite{Qi2019}). Finally, we note from the top row of Fig.~\ref{fig:wheatallsubs1}--\ref{fig:teflonallsubs1} that the contact angle dependence ($\theta_{\text{L}}$ versus $t$ and $\theta_{\text{R}}$ versus $t$) and the droplet height dependence ($H$ versus $t$) are all concave downwards in these examples.

The situation is more complicated for a high concentration of surfactant, 1\% S465 (bottom row of Fig.~\ref{fig:wheatallsubs1} and \ref{fig:teflonallsubs1}).  Here we see that evaporation on Teflon is qualitatively very similar to both wheat and Teflon with 0.1\% S465 as just described (that is, CCR mode for the entire time with concave downward contact angle and droplet height dependence on time).  In contrast, evaporation on wheat with 1\% S465 appears to be very different.  In this case (bottom row of Fig.~\ref{fig:wheatallsubs1}), the contact radius $R$ is not constant and so the mode of evaporation is not CCR; instead, the contact radius appears to increase for the first part of the experiment and then decreases down to zero in a seemingly haphazard manner in what may be considered a type of stick-slip mode.  In another difference, both the contact angle dependence and the droplet height dependence ($H$ versus $t$) are all concave upwards.  Finally, it is clear that the total evaporation time for 1\% S465 on wheat is much shorter than the other cases of wheat and Teflon.

\begin{figure*} [h!]
	\centering
	\includegraphics[width=0.6\textheight,keepaspectratio]{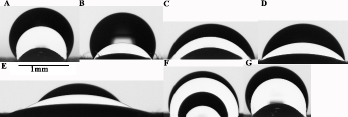}
	\caption{Time lapse images. The white droplets have had the colours inverted. The first row is horizontal droplets and the second row is special cases; E is horizontal and F and G are 90$\,^{\circ}$ incline. All solutions are 0.1\% S465 except E, which is 1\% S465 and F, which is water. The surfaces are Teflon (A and G), wheat (B and E), Parafilm (C) and capsicum (D and F). The times in secs for A is 0, 840, 1380; B is 0, 1260, 1530; C is 0, 600, 1080; D is 0, 900, 1320; E is 0, 30, 270; F is 0, 300, 990, 1470 and G is 0, 420, 1050. Scale bar is 1mm and is the same for all images. In A, B, D and G we can see the CCR mode. In C and F we can see the CCR then CCA modes. In E we see the spreading that occurs only on wheat with 1\% S465. In G we can see the large difference in the left and right contact angles due to the 90$\,^{\circ}$ incline and the difference decreases with time.}
	\label{fig:alldrops1}
\end{figure*}

In order to visualise the evaporation modes and the droplet shapes as the (spreading and) evaporation processes take place, we have included in Fig.~\ref{fig:alldrops1} examples of images of droplets shapes in our experiments.  In Fig.~\ref{fig:alldrops1}(A) and (B) we show images of 0.1\% S465 on Teflon and wheat, respectively; here the CCR mode is clearly evident.  Further, for these two examples on horizontal surfaces, the spherical cap nature of the droplets is illustrated.  As a contrast, in (E) we show 1\% S465 on a horizontal wheat surface where it is clear that the droplet radius is not fixed in time.

\begin{figure*}  [h!]
	\centering	
\includegraphics[height=0.3\textheight,keepaspectratio]{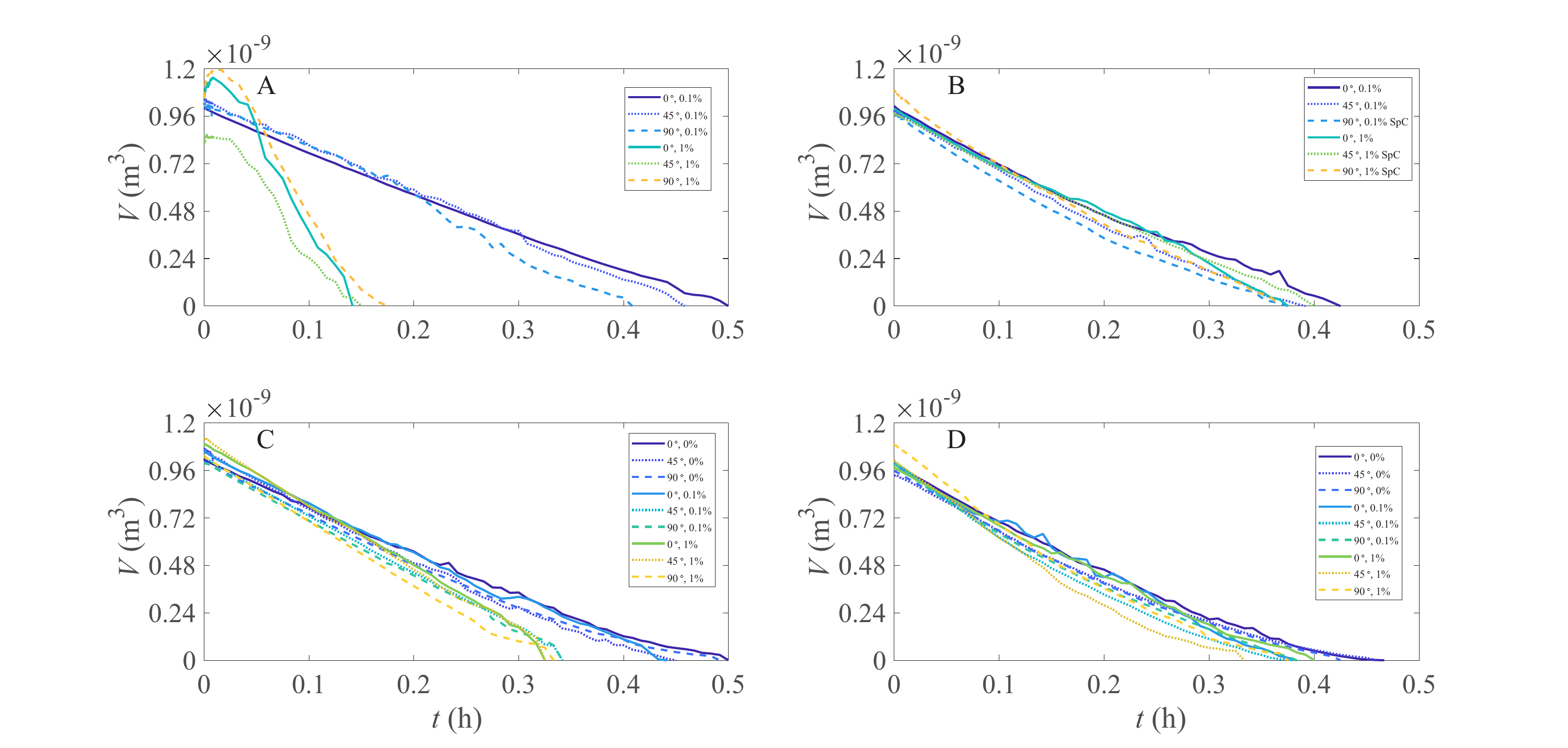}
	\caption{Experimental data for droplet evaporation on wheat (A), Teflon (B), capsicum (C) and Parafilm (D), for droplet volume (m$^3$) with time (h). The surface incline and solution are shown in the legend. For (B), SpC indicates where the spherical cap volume equation (see equation (S1) in the Supplementary Material) on an incline is used instead of the volume calculated within the experimental apparatus.}
	\label{fig:teflonallvolume1}
\end{figure*}

Measurements of the time-dependent change in droplet volume are provided in Fig.~\ref{fig:teflonallvolume1}.
Wheat and Teflon are shown in (A) and (B), respectively.  It is clear that the increase in surfactant concentration from 0.1\% to 1\% S465 has a dramatic effect on the volume and evaporation rate for wheat but no significant effect on Teflon.  Note there were some challenges with calculating droplet volume with the CAM instrument software, as discussed in the Supporting Information. We shall return to Fig.~\ref{fig:alldrops1} and Fig.~\ref{fig:teflonallvolume1} later to discuss the results on capsicum and Parafilm.

With all these observations in mind, we draw the following conclusions about evaporation on wheat and Teflon.  Provided the concentration of surfactant is low enough, the evaporation process appears to be routine and should be amenable to a mathematical model that assumes a CCR mode.  Further, at these low concentrations, Teflon acts as a synthetic analogue of wheat for the purposes of evaporation.  In part, this close connection between the two substrates is not surprising as we have chosen to use Teflon due to its similar surface properties to wheat.  On the other hand, for sufficiently high concentration of surfactant, it is clear that while the evaporation process on Teflon continues to be routine, evaporation on wheat is nonstandard.  Therefore, we certainly cannot consider Teflon as a synthetic analogue of wheat at high surfactant concentrations.

It is worth discussing these differences in more detail and speculating on the mechanisms causing the nonstandard evaporation profile of 1\% S465 on wheat.  Teflon is significantly rougher and more non-polar than wheat. Even with the much higher surfactant concentration, the required energy barrier for significant spreading has not been overcome.  As such, the contact angle for Teflon with 1\% S465 is only slightly less than that for 0.1\% S465 and certainly remains well above 90$\,^{\circ}$.  In contrast, the increase in surfactant concentration from 0.1\% to 1\% on wheat is sufficient to overcome this energy barrier, thereby significantly decreasing the contact angle and allowing the droplet to greatly wet the surface.

Further, studies have shown that adjuvants can adsorb to leaves \cite{Zhang2017Soft,Zhang2006}, S465 can adsorb to surfaces \cite{Musselman2002adsorption} and adsorption of non-ionic surfactants on hydrophobic surfaces may lead to a complete replacement of the surface by the adsorbed surfactant as a monolayer, affecting the surface mechanisms \cite{Zhang2006}. Therefore, the nonstandard evaporation profile of 1\% S465 on wheat could be due to surfactant adsorption and the surface properties of wheat.  The dynamics of evaporation of 1\% S465 on wheat leaves shares some similarities with that reported by Peirce et al.~\cite{Peirce2016}, who consider initial spreading and subsequent evaporation of surfactant droplets on wheat, surmising that this regime is caused by adsorption of surfactant.

Another possible cause of the nonstandard evaporation behaviour of 1\% S465 on wheat could be due to the shape of the droplet.  By visualising the droplets attached to the surfaces, as shown in Fig.~\ref{fig:allsurf902}, we can see that the droplets with 1\% S465 on wheat are not well approximated by a spherical cap and did not have a near circular contact line. This figure shows images of water, 0.1\% and 1\% S465 droplets on a 90$\,^{\circ}$ incline, on the upper leaf surfaces of wheat (A,B) and capsicum (D), Teflon (C) and Parafilm (E).   In Fig.~\ref{fig:allsurf902} (A) and (B), the 1\% S465 droplet has spread much further than the 0.1\% S465 droplet and has a much lower contact angle and larger contact area. It is difficult to see the exact shape of the contact line of the droplet with 1\% S465, however it appears to be roughly elliptic, where the longitudinal edges of the droplet slip into the deep ridges of the wheat leaf. The other droplets in Fig.~\ref{fig:allsurf902} appear to be closer to spherical caps. In Fig.~\ref{fig:allsurf902} (B) and (C), we see the 0.1\% S465 droplet (left) on wheat and Teflon appear similar, while the 1\% S465 droplet (right) on wheat and Teflon are clearly a different shape.  This elongated droplet shape (right-most image in Fig.~\ref{fig:allsurf902}(A,B)) that appears only for 1\% S465 on wheat is likely to contribute significantly to the much faster evaporation rate shown in Fig.~\ref{fig:teflonallvolume1}(A).  This reasoning is consistent with the findings of Jansen et al.~\cite{Jansen2015}, who set up a striped patterned substrate that is similar in some respects to wheat which had the effect of stretching the droplets out in the direction parallel with the stripes \cite{Bliznyuk2009}. These elongated droplets were shown to evaporate much faster than those which take the shape of a spherical cap \cite{Jansen2015}, with the explanation being that the elongated droplets have a larger surface area in contact with the air and therefore increase the rate of vapour diffusion.

\subsection{Experiments on capsicum and Parafilm}  \label{sec:expcapparafilm}

\begin{figure*} [h!]
	\includegraphics[width=0.8\textwidth, center,keepaspectratio]{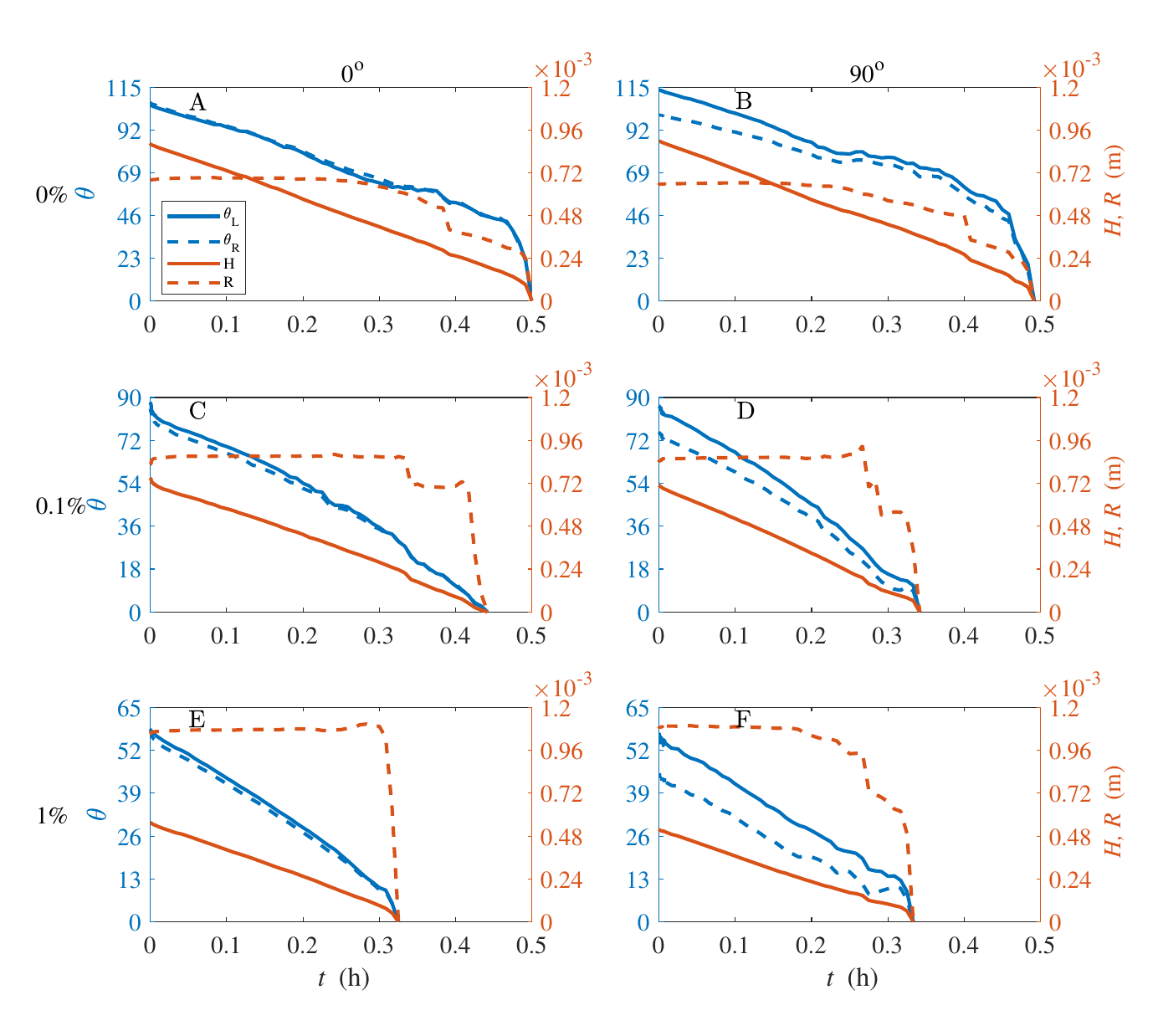}
	\caption{Experimental data for droplet evaporation on capsicum leaves. Water (0\% S465)  is shown in A and B; 0.1\% S465 is shown in C and D; and 1\% S465 is shown in E and F. The columns of plots correspond to 0$\,^{\circ}$ and 90$\,^{\circ}$ surface incline. The left and right contact angles, $\theta_L$ and $\theta_R$, are in blue, corresponding to the left hand side axis, while the contact radius $R$ and height $H$ are both in red and correspond to the right hand side axis.  The horizontal axis measures time, $t$.}
	\label{fig:capcicumallsubplots1}
\end{figure*}

\begin{figure*} [h!]
	\includegraphics[width=0.8\textwidth, center,keepaspectratio]{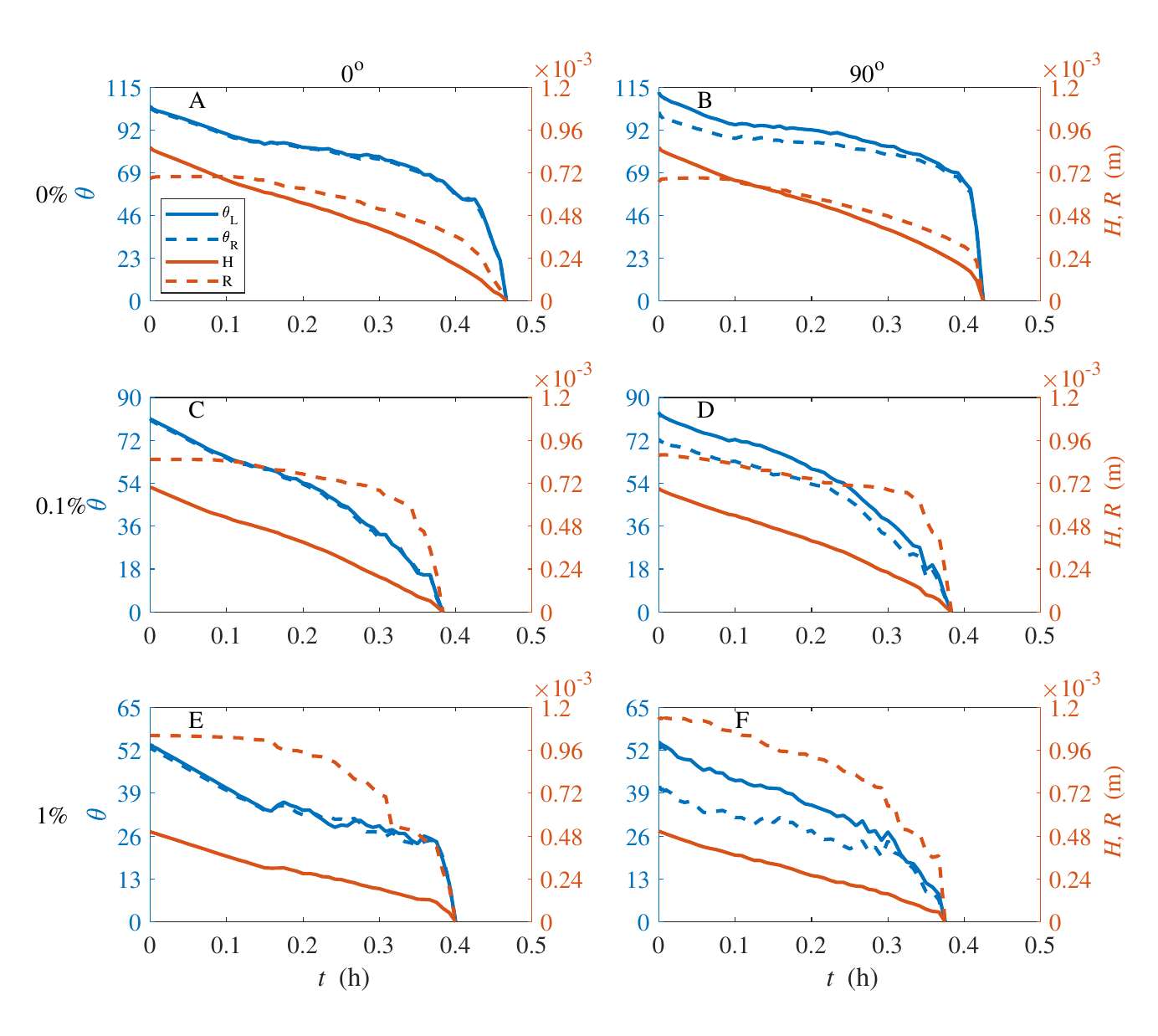}
	\caption{Experimental data for droplet evaporation on Parafilm. Water (0\% S465) is shown in A and B; 0.1\% S465 is shown in C and D; and 1\% S465 is shown in E and F. The columns of plots correspond to 0$\,^{\circ}$ and 90$\,^{\circ}$ surface incline. The left and right contact angles, $\theta_L$ and $\theta_R$, are in blue, corresponding to the left hand side axis, while the contact radius $R$ and height $H$ are both in red and correspond to the right hand side axis.  The horizontal axis measures time, $t$.}
	\label{fig:Parafilmallsubplots1}
\end{figure*}

We now turn to our results for evaporation on capsicum and Parafilm, recalling that both capsicum and Parafilm have non-polar surfaces that are smooth and either moderate-to-wet or difficult-to-wet.  We begin with Fig.~\ref{fig:capcicumallsubplots1} and \ref{fig:Parafilmallsubplots1} which show plots of the left and right contact angles, the contact radius $R$ and droplet height $H$ versus time.  In these figures we have included water in the results (top row) since water droplets were able to adhere on these surfaces (as opposed to our wheat and Teflon, on which water would not adhere).  We can see that the evaporation profiles are similar for capsicum and Parafilm, indicating that Parafilm acts as a reasonable synthetic analogue for capsicum in these cases.  S465 has a significant impact on the initial contact angle on both capsicum and Parafilm; for example, on horizontal capsicum the initial contact angles for water, 0.1\% and 1\% S465 are 105$\,^{\circ}$, 85$\,^{\circ}$, 58$\,^{\circ}$, respectively.  Further, S465 has a strong effect on the initial droplet spread area, as expected, which can be seen by noting the red dashed curves that denote the droplet radius $R$ in Fig.~\ref{fig:capcicumallsubplots1} and \ref{fig:Parafilmallsubplots1}.  This reduction in $\theta_0$ and increase initial droplet radius with increasing concentration of surfactant leads to more rapid evaporation.

The key observation from Fig.~\ref{fig:capcicumallsubplots1} and \ref{fig:Parafilmallsubplots1} is that for most of these examples on capsicum, the evaporation mode is CCR for much of the time, sometimes followed by a period of time characterised by a combined mode (C) where both the contact angle and radius are decreasing.  For Parafilm the combined mode is more prevalent (for all examples, we have classified the mode of evaporation in Table S1).

The combined mode observed here on capsicum and Parafilm is not commonly described in the literature. As the combined mode appears to occur in our experiments with pure water, adsorption of S465 is not likely to be a contributing factor. Importantly, it is clear from observing the videos captured during the experiments that the combined mode involves one side of the contact line pinned to the substrate while the other side of the droplet radius shrinks for a few minutes, then the side that is pinned can switch and this behaviour continues until extinction.  Given that capsicum and Parafilm are smooth surfaces, this type of irregular behaviour is reminiscent of the `snap evaporation' studied recently by Wells et al.~\cite{Wells2018} on smooth but wavy surfaces.  This partial de-pinning behaviour may contribute to the trend in the contact angle and does not occur with wheat and Teflon, even when the initial contact angles are similar.  Therefore we conclude that the surface characteristics of capsicum and Parafilm have properties that allow the partial de-pinning of the droplet. \citet{Lam2002} found a combined evaporation mode can occur when using an $n$-pentadecane droplet on FC-732-coated silicon wafer and note this behaviour could be due to a strong interaction such as a chemical bond.

Returning to Fig.~\ref{fig:alldrops1}, where we present time lapse images of evaporating droplets, we can easily see the effect of the much smaller contact angle that is present on both capsicum (D) and Parafilm (C) when compared to Teflon (A) and wheat (B).  Further, while evaporation of 0.1\% S465 on Teflon and wheat is clearly in CCR in (A) and (B), the equivalent experiment on Parafilm and capsicum is not so clear (C and D). Indeed, while these images lend support to the observation that capsicum (D) is evaporating under CCR with 0.1\% S465, the evaporation mode for Parafilm (C) appears from these images to be mixed.

The dependence of droplet volume on time is presented in Fig.~\ref{fig:teflonallvolume1} for capsicum (C) and Parafilm (D).  For these two surfaces, it is remarkable how the overall rate of volume loss is only weakly affected by both surfactant and the angle of incline, at least when compared to those effects on wheat (A) and Teflon (B).  Another point worth noting with all of our experimental results relates to the dependence of evaporation time on inclination angle.  In their study of evaporation of water droplets on an incline, \citet{Kim2017} found that the total evaporation time for large (8~$\mu$L) droplets increased as inclination increased to $90\,^{\circ}$, in contrast to our observations for moderately large (1~$\mu$L) droplets (which is at the very upper end of the droplet spectrum used in agrochemical spraying).  Therefore, the results of \citet{Kim2017} are not likely to be relevant for agricultural spray applications.

\subsection{Mathematical modelling results}  \label{Model_results}

\begin{figure*} [h!]
\centering
	\includegraphics[height=0.16\textwidth,keepaspectratio]{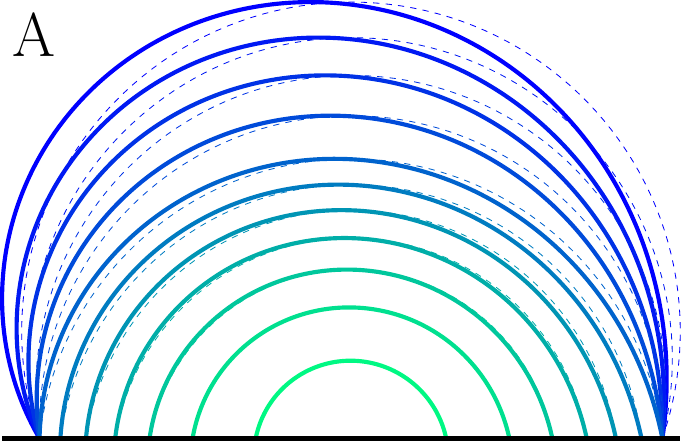}\hspace{3ex}
	\includegraphics[height=0.16\textwidth,keepaspectratio]{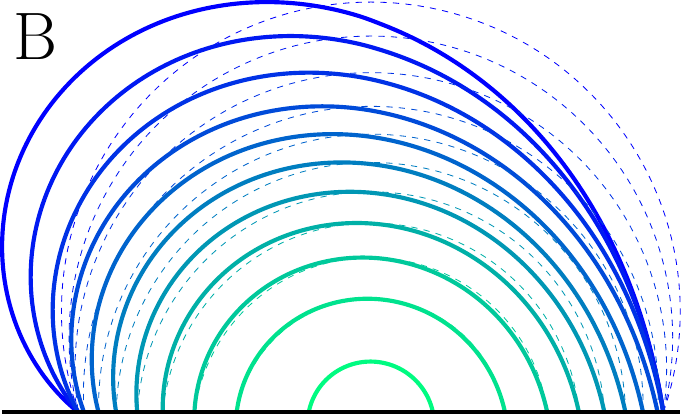} \\

\vspace{3ex}\hspace{-8ex}\includegraphics[height=0.19\textwidth,keepaspectratio]{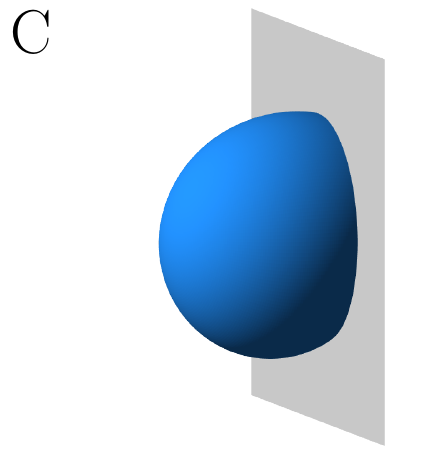}\hspace{8ex}
	\includegraphics[height=0.19\textwidth,keepaspectratio]{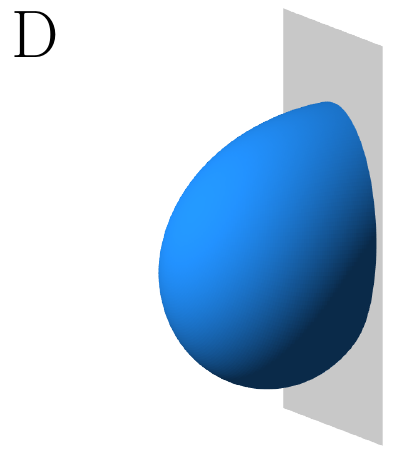}
\caption{(A)-(B) Illustrative numerical simulations of the Young-Laplace-Popov model with droplet profiles (thick solid curves) drawn for equally spaced times.  The surface tension used in the simulations is the same as a solution with 1\% S465.  The substrate is inclined at 90$\,^{\circ}$ so that gravity points to the left in these images.  Initially, the droplet's volume and height are chosen to be (A) 1~$\mu$L and $0.91$~mm (B) 4~$\mu$L and $1.45$~mm.  As a comparison only, semi-circles are drawn (thin dashed curves) to compare the droplet shape with a spherical cap.  (C)-(D) Three-dimensional images of the initial droplet shapes, computed with the Young-Laplace equation.  (C) shows the initial droplet from (A), while (D) shows the initial droplet from (B).  Note that the figures are not all drawn on the same scale; in reality, the droplets in (A) and (C) are much smaller than the droplets in (B) and (D).}
	\label{fig:simulations}
\end{figure*}

We begin this section by briefly demonstrating how droplets evolve according to the YLP model outlined in section~\ref{sec:numericalMethod}.  In Fig.~\ref{fig:simulations} we present illustrative simulations for sessile droplets evaporating on a 90$\,^{\circ}$ incline according to the YLP model.  The droplet profiles (the thick solid curves) in Fig.~\ref{fig:simulations}(A) are for a droplet whose initial volume is 1~$\mu$L and initial droplet height is $H_0=0.91$ mm, while the profiles in (B) are for an initially 4~$\mu$L droplet with $H_0=1.45$ mm.  Using this information, the model solves the Young-Laplace equation to generate the initial droplet shape (also shown in Fig.~\ref{fig:simulations}(C)-(D)), which has contact angles (A) $\theta_{\text{L}0}=120.3^\circ$, $\theta_{\text{R}0}=98.2^\circ$, (B) $\theta_{\text{L}0}=140.9^\circ$, $\theta_{\text{L}0}=81.1^\circ$, corresponding to a hypothetical superhydrophobic surface.  The surface tension used in the simulation is for 1\% S465, which corresponds to a capillary length of roughly $\lambda_c=\sqrt{\gamma/(\Delta\rho g)}=1.7$ mm (slightly less than the capillary length for water, which is $2.7$ mm).  In the simulations, the evaporation mode is initially CCR until the right contact angle reaches an arbitrary threshold value of $\theta_{\text{rec}}=78^\circ$, after which point the evaporation is assumed to follow the CCA mode.  The total evaporation times for these simulations are (A) 1550 sec and (B) 3970 sec.

Using $H_0$ as the representative length scale, the Bond numbers in Fig.~\ref{fig:simulations}(A) and (B) are $\mathrm{Bo}\,= (H_0/\lambda_c)^2=0.29$ and $0.73$, respectively.  We see here that for droplets of these sizes, the Bond numbers are roughly of order one, which means that gravity is having a significant, but not an overwhelming, effect.  For comparison, we have included circular arcs (thin dashed curves) which represent the shape the droplets would take if the radius and height were fixed but gravity was neglected.  Clearly, the larger of the two initial droplets has the higher Bond number and the most distorted shape.  As the droplet evaporates, the droplet height will inevitably reduce to become much smaller than the capillary length; for these later times the droplet is essentially a spherical cap and so gravity is negligible.  Indeed, in Fig.~\ref{fig:simulations} we can see for later times the droplet profiles (thick solid curves) and the circular arcs (thin dashed curves) are visually indistinguishable.  Note that, when compared with droplets on a horizontal surface, droplets on an incline have a larger surface area and therefore evaporate faster.

It is worth remembering that simpler model in Sections~\ref{sec:Popov} is not able to compute actual droplet shapes like those presented in Fig.~\ref{fig:simulations}.  Instead, for evaporation on an incline, to employ the Popov model from Section~\ref{sec:Popov} we simply average the left and right contact angles.  So while the obvious advantage of the YLP model (Section~\ref{sec:numericalMethod}) is that it incorporates the effects of surface tension and gravity on the droplet shape in the simulations, the advantage of the simpler model is that it does not require the numerical solution of a partial differential equation in a complex domain.

\begin{figure*} [h!]
	\includegraphics[width=1\textwidth,keepaspectratio]{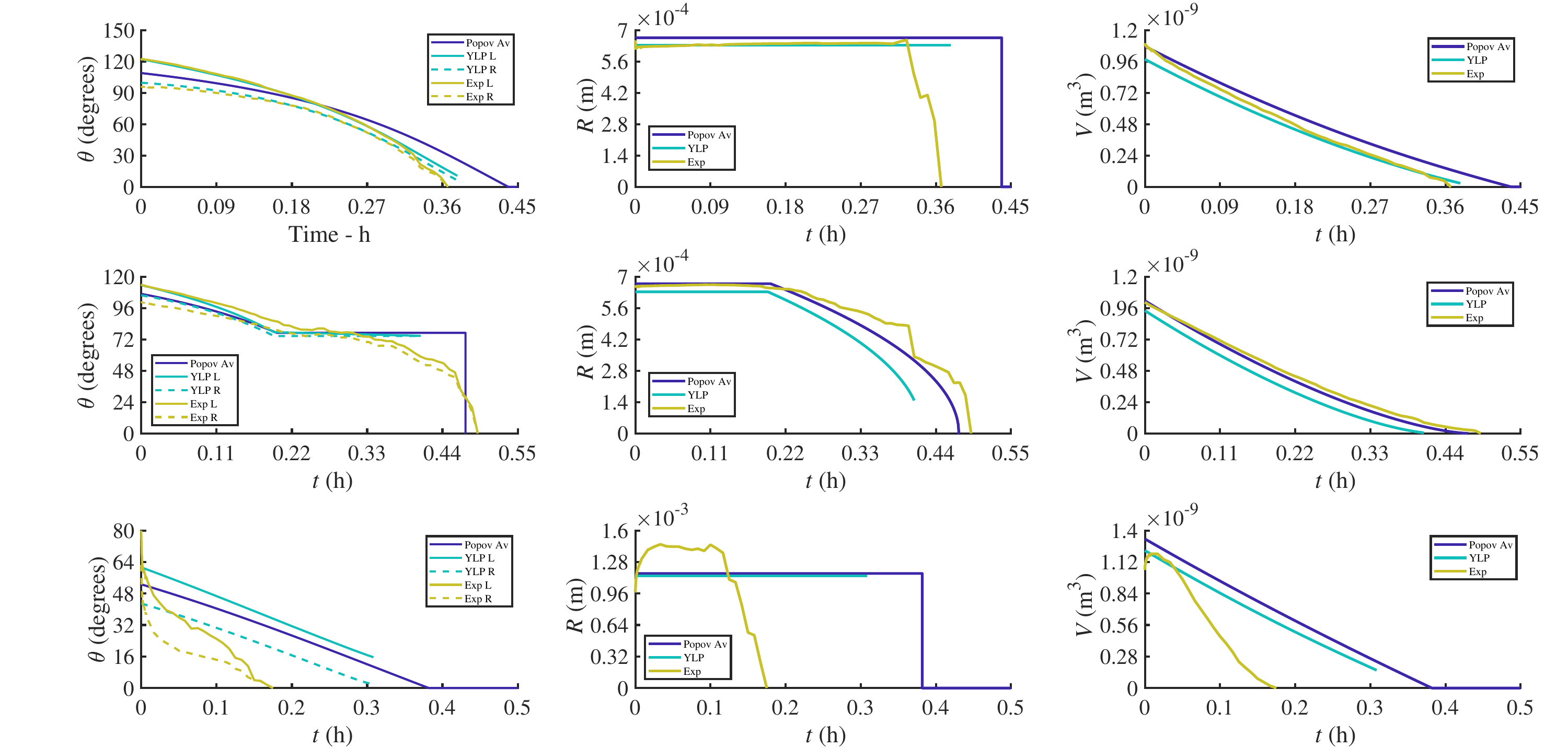}
\caption{Numerical simulations of the three models compared with experimental results on a 90$\,^{\circ}$ incline.  The first row is for Teflon with 1\% S465, the second row is for capsicum with water, while the third row is for wheat with 1\% S465.  In these plots, Popov Av indicates the Popov model with the left and right initial contact angles averaged, YLP is the Young-Laplace-Popov model and Exp is the experimental data.}
	\label{fig:modellingfigure}
\end{figure*}

Returning to the experimental data,
Fig.~\ref{fig:modellingfigure} shows a selection of the modelling results compared to the experimental data, all for the case of a 90$\,^{\circ}$ incline (additional results are shown in Fig.~S16--S30 in the Supplementary Material).  For these figures, YLP is the Young-Laplace-Popov model described by (\ref{Laplace1})--(\ref{eq:Vflux}) (using the larger contact angle $\theta_{\text{L0}}$ and droplet height $H_0$ as the required inputs to compute the initial shape), Popov Av is the \citet{Popov2005} model described by (S2)--(S4) (with the left and right initial contact angles averaged to find $\theta_0$), and Exp is the experimental data.  The contact angles $\theta$ are shown for each side, left (L) and right (R), together with the contact radius $R$ and droplet volume $V$, all plotted against time $t$.

In the first row of Fig.~\ref{fig:modellingfigure}, which is for Teflon with 1\% S465, the models are run in CCR mode for the entire evaporation time, while for the second row, which is for capsicum with water, the process is assumed to be in CCR mode until $t=t_{\text{rec}}$, at which point it transitions to the CCA mode.  In each of these three figures, the YLP compares well to the experimental data. The Popov Av model, considering its simplicity, is also a quite a good representation of the experimental data. As the experimental inputs to the model are the initial height and the contact angles (for the YLP model we only require the initial value of the larger contact angle $\theta_{\text{L}}$), while the initial contact radius and volume are calculated, it is expected that the contact angles and droplet height (see Figs. (S16)--(S30)) compare the most reliably to the experimental data over time, and this is what we see in our results.%

For the cases with capsicum and Parafilm in which we have assumed a CCR/CCA mode in the models (such as the second row of Fig.~\ref{fig:modellingfigure}), the predicted outputs for the droplet height, contact radius and volume compare reasonably well with the data, however the contact angles towards the end of the droplet lifetime do not compare well. As observed in the example in the second row of Fig.~\ref{fig:modellingfigure} (and summarised in Table~S1), the main reason for this discrepancy is that many of these experimental results actually demonstrate an approximate CCR mode followed by a combined mode (C), but our models cannot handle a combined mode.  Instead, we choose to simulate any observed combined mode as a CCA model as the contact angles are changing slower than the contact radius.  Developing and studying a mathematical model for a combined mode on an incline, where both the contact angles and radius change simultaneously, is certainly of interest but would require a much more sophisticated set of governing equations and/or a treatment via a computational fluid dynamics (CFD) framework.

A notable case in terms of the experimental data is evaporation on wheat with 1\% S465 (see the second row of Fig.~\ref{fig:wheatallsubs1}). As highlighted in Section~\ref{sec:expwheatteflon}, the evaporation profile of 1\% S465 on wheat exhibits S/CCR/C evaporation modes and appears very different from the case with 0.1\% S465.  For 1\% S465 on wheat, we have in the third row of Fig.~\ref{fig:modellingfigure} compared the predictions from our models with the experimental data in the most extreme case of a 90$\,^{\circ}$ incline.  This comparison is understandably poor, as the models do not allow for S/CCR/C evaporation modes.  In any event, we have attempted to begin the models after the spreading regime has ended and simulated the results with the CCR mode alone. We can see the droplet evaporates very quickly, and the models over predict the evaporation times by a significant margin.  It is clear the simple models we use in this paper are not able to accurately recover the experimental data (for this case of 1\% S465 on wheat), indicating additional mechanisms are required in the models.   One way to adapt our YLP model is to allow for a non-circular contact line, for example initially an ellipse.  This particular change is possible, however in practice we would require further information from our experimental approach (for example, the contact angles and radius from the minor axis of the droplet) in order to fully specify the problem.  Another such mechanism is the adsorption of surfactant onto the leaf surface, as we discuss in Section~\ref{sec:expwheatteflon}.  Further, it is clear from Fig.~\ref{fig:allsurf902}(B) (right photograph) that the 1\% S465 droplet on wheat at 90$\,^{\circ}$ does not have a circular contact line, so that constraint could be dropped in a future model.  To motivate one of many other possibilities, a study~\cite{DEFRAEYE2013} found that small variations in stomatal density had a significant effect on droplet evaporation. As the droplet coverage area is the largest with 1\% S465 on wheat of all our results, hence covering more stomata, the effect of stomata may need to be included in a future evaporation model.

We close this section by noting that the authors are working on a larger project aiming to build interactive software for simulating agrochemical spraying of crops \cite{Dorr2016118,Zabkiewicz20}. In terms of future research, it would be useful to incorporate information from the present study into this software.

\section{Conclusions}
The aims of this study were to present new experimental data on evaporation of surfactant droplets on horizontal and inclined leaves, to investigate whether these evaporation processes on leaves are analogous to those on synthetic surfaces with similar surface properties, and to present new simplified mathematical models to simulate the experiments.

The novel experimental data in this paper consists of droplet evaporation with time on wheat, Teflon, capsicum and Parafilm surfaces with water, 0.1\% and 1\% S465 on a 0$\,^{\circ}$, 45$\,^{\circ}$ and 90$\,^{\circ}$ incline.
The effect of increasing the concentration of S465 is significant, especially on wheat, and increases the rate of evaporation with decreased initial contact angle and increased droplet spread area. The most notable case is 1\% S465 on wheat which is a clear outlier; here, the droplet appears elongated with the major axis running parallel with the ridges along the wheat leaf surface, in some respects similar to experimental results produced with striped patterned substrate \cite{Jansen2015}, where evaporation rates were also significantly increased due to this geometry.  The degree to which these observations on wheat can be explained using fabricated anisotropic surfaces is worth further study \cite{Xia2012}.   On the other hand, the effect of surface inclination angle on evaporation is dependant on surface properties and surfactant concentration, but appears to be less influential compared to the effect of increasing the surfactant concentration. Indeed, for many of our examples, low concentration of surfactant on wheat and Teflon, we observed the constant contact radius mode for the lifetime of the evaporation even on a 90$\,^{\circ}$ incline, which illustrates how our very difficult-to-wet surfaces can strongly pin the contact line in a similar manner to that demonstrated very recently \cite{Qi2019} on vertical surfaces (that were biphobic by construction).

In terms of comparing evaporation on leaf surfaces to our synthetic surfaces, we have chosen to pair  wheat with Teflon as they are both rough, non-polar and very difficult-to-wet, and pair capsicum with Parafilm as they are both smooth and either moderate-to-wet or difficult-to-wet.  We conclude that for low surfactant concentrations Teflon can be interpreted as a synthetic analogue for wheat, while for all concentrations Parafilm is a reasonable analogue for capsicum.  On the other hand, for high surfactant concentrations, evaporation on wheat is nonstandard and very different to that on Teflon; therefore, in this regime we cannot consider Teflon as a synthetic analogue of wheat.

It is worth emphasising that the qualitative differences we observe for 1\% S465 on wheat when compared to the other case cannot be explained solely by contact angle effects.  In particular, we note that with 1\% S465, the initial contact angle on wheat (on a horizontal surface) is roughly 70$^\circ$ (55$^\circ$ after the short spreading phase), which is much lower than the roughly 130$^\circ$ which we see with 0.1\% S465 on wheat.  Therefore the increase in surfactant concentration has obviously greatly increased the droplets' ability to wet the wheat surface.  However, with 1\% S465, the initial contact angle on capsicum is roughly 59$^\circ$, which is very close to 55$^\circ$, and the 1\% S465 on capsicum case appears to be essentially following the same trends as the other cases considered in this study.  Therefore, we conclude that the differences between 1\% S465 on wheat and the other cases cannot be simply due to the ability of the 1\% S465 droplet being able to wet wheat more (or else we would see some evidence of these new behaviours in the case of 1\% S465 on capsicum).  Instead, the effects of surface chemistry, roughness and anisotropy are playing important roles, which is why wetting phenomena on leaf surfaces is so fascinating.

We have applied two mathematical models to simulate our experiments.  The first is what we call the Popov model, which is a well-known quasi-steady diffusion limited model that applies on a horizontal surface.  This model assumes the droplet shape is a spherical cap and that evaporation is taking place in the constant contact radius model followed by the constant contact angle mode, or simply the constant contact radius mode alone.  For evaporation on an inclined surface, in order to implement the Popov model we make an adjustment by simply taking the average of the largest and smallest contact angles as an input to the model. The second is our new Young-Laplace-Popov, which builds on the Popov model by allowing for the droplet to take a natural shape due to the influence of gravity, surface tension and the incline.  A feature of these models is that they are relatively easy to simulate and do not require the considerable computational set-up and expense of typical computational fluid dynamics approaches \cite{Saenz2017dynamics}.

The models can predict evaporation in most cases on an incline with minimal inputs of left and right initial contact angle and height, however there are two cases in which the models are less accurate.  The first case is the combined mode on capsicum and Parafilm, where both the contact radius and contact angle changes in time simultaneously.  We have found that the models are still reasonable if we approximate the combined mode with a constant contact angle mode.  The second case is 1\% S465 on wheat, where the models are not at all able to predict the experimental results.  In the experiments the droplets evaporate very quickly, but the models over-predict the total evaporation time by a significant margin. It is hypothesised that additional mechanisms required by the models for this case of high surfactant concentration on wheat leaves may include adsorption of S465 to the wheat leaf surface, stomatal effects on evaporation or, most likely, a contact line that is not circular but instead is determined by the complex interaction between the droplet and the ridges that run along the surface of wheat leaves.

\section*{Conflict of interest}
There are no conflicts to declare.
\section*{Acknowledgements}
Experimental components were performed at PPC$_{\text{NZ}}$, Rotorua, New Zealand. The authors gratefully acknowledge support from the Australian Research Council through the ARC Linkage Project (LP160100707) and the associated industry partners Syngenta and Nufarm.

\section*{Appendix A.~Supplementary Material}
Supplementary data associated with this article can be found in the document ``Supplementary Material for Evaporating Droplets on Inclined Plant Leaves and Synthetic Surfaces: Experiments and Mathematical Models''.
%

\end{document}